\documentclass[twocolumn,english,aps,prb,floats]{revtex4}
\usepackage[T1]{fontenc}
\usepackage[latin9]{inputenc}
\usepackage{amsmath}
\usepackage{amssymb}
\usepackage{graphicx}
\usepackage{esint}

\makeatletter

\@ifundefined{textcolor}{}
{%
 \definecolor{BLACK}{gray}{0}
 \definecolor{WHITE}{gray}{1}
 \definecolor{RED}{rgb}{1,0,0}
 \definecolor{GREEN}{rgb}{0,1,0}
 \definecolor{BLUE}{rgb}{0,0,1}
 \definecolor{CYAN}{cmyk}{1,0,0,0}
 \definecolor{MAGENTA}{cmyk}{0,1,0,0}
 \definecolor{YELLOW}{cmyk}{0,0,1,0}
}

\@ifundefined{textcolor}{}{%
 \definecolor{BLACK}{gray}{0}
 \definecolor{WHITE}{gray}{1}
 \definecolor{RED}{rgb}{1,0,0}
 \definecolor{GREEN}{rgb}{0,1,0}
 \definecolor{BLUE}{rgb}{0,0,1}
 \definecolor{CYAN}{cmyk}{1,0,0,0}
 \definecolor{MAGENTA}{cmyk}{0,1,0,0}
 \definecolor{YELLOW}{cmyk}{0,0,1,0}
}

\usepackage{babel}
\usepackage{bm}

\usepackage{babel}

\usepackage{babel}

\makeatother

\usepackage{babel}
\begin{document}

\title{Thermal Creation of Skyrmions in Ferromagnetic Films with Perpendicular
Anisotropy and Dzyaloshinskii-Moriya Interaction}

\author{Dmitry A. Garanin$^{1}$, Eugene M. Chudnovsky$^{1}$, Senfu Zhang$^{2}$, and Xixiang
Zhang$^{2}$}

\affiliation{$^{1}$Physics Department, Herbert H. Lehman College and Graduate
School, The City University of New York, 250 Bedford Park Boulevard
West, Bronx, New York 10468-1589, USA \\
 $^{2}$Physical Science and Engineering Division (PSE), King Abdullah
University of Science and Technology (KAUST), Thuwal 23955-6900, Saudi
Arabia}

\date{\today}
\begin{abstract}
We study theoretically, via Monte Carlo simulations on 
lattices containing up to $1000\times1000$ spins, thermal creation of skyrmion lattices in a 2D ferromagnetic
film with perpendicular magnetic anisotropy and Dzyaloshinskii-Moriya
interaction. At zero temperature, skyrmions only appear in the magnetization
process in the presence of static disorder. Thermal fluctuations violate
conservation of the topological charge and reduce the
effective magnetic anisotropy that tends to suppress skyrmions. In accordance with recent experiments, we find that elevated
temperatures assist the formation of skyrmion structures. Once such a structure is formed,
it can be frozen into a regular skyrmion lattice by reducing the temperature.
We investigate topological properties of skyrmion
structures and find the average skyrmion size. Energies of domain
and skyrmion states are computed. It is shown that skyrmion lattices have lower energy than labyrinth domains within a narrow field range.  
\end{abstract}
\maketitle

\section{Introduction}

Skyrmions are considered for applications in spintronic devices and
information processing due to their topological stability \cite{Nagaosa2013,Tomasello,Zhang2015,Klaui2016,Leonov-NJP2016,Hoffmann-PhysRep2017,Fert-Nature2017}.
Initially introduced in nuclear physics \cite{SkyrmePRC58,Polyakov-book},
skyrmions entered magnetism after it was realized \cite{BelPolJETP75}
that the simplest exchange model in two dimensions (2D) describes
topological defects in ferro- and antiferromagnets \cite{Lectures,Brown-book}.
During the last decade, magnetic skyrmions have been widely observed
in slabs \cite{Grigoriev2009,Pappas2009,Jonietz2010,Munzer2010,Yu2010,Kanazawa2011,Yu2011,Tonomura2012,Yu2012,Milde2013,Shibata2013,Park2014}
and multilayered films \cite{Schlenhoff2015,Boulle2016,Moreau-2016,Woo2016,Upad2016,Legrand2017,Pollard2017,Soum2017}.

A pure continuous-field 2D exchange model is scale invariant, making
the energy of the skyrmion independent of its size. In a solid, the
scale invariance is inevitably violated by the crystal lattice and
various interactions other than the exchange. Consequently, the energy
of the skyrmion becomes dependent on its size. The crystal lattice
alone makes the energy of the skyrmion go down with decreasing size,
leading to skyrmion collapse \cite{CCG-PRB2012}. Perpendicular magnetic
anisotropy, dipole-dipole interaction, magnetic field, and confined
geometry can stabilize significantly large magnetic bubbles with skyrmion
topology \cite{MS-bubbles,ODell,Moutafis-PRB2009,Ezawa-PRL2010,Makhfudz-PRL2012,Montoya2017,GCZ-EPL2017}.
The stability of small skyrmions requires other than Heisenberg exchange
coupling \cite{Chen-APL2015,Lin-PRB2016,Rozsa2016,Malottki2017,Han},
more complex magnetic anisotropy \cite{AbanovPRB98,IvanovPRB06,IvanovPRB09},
static disorder \cite{EC-DG-PRL2018,EC-DG-NJP2018}, or
a non-centrosymmetric system with large Dzyaloshinskii-Moriya interaction
\cite{Dzyaloshinskii,Moriya,Bogdanov-Nature2006,Heinze-Nature2011,Leonov-NatCom2015,Leonov-NJP2016}.

When working with skyrmions, one of the most challenging tasks is
creation of skyrmions in a predictable manner. Creating, annihilating
and moving skyrmions by current-induced spin-orbit torques has been
the most commonly used method so far, see, e.g., Refs.\ \onlinecite{Yu-NanoLet2016,Fert-Nature2017,Legrand2017,He-2017}.
Skyrmion bubbles have been generated by pushing elongated magnetic
domains through a constriction using an in-plane current \cite{Jiang-Sci2015,Hoffmann-PhysRep2017}.
Small skyrmions can be also written and deleted in a controlled fashion
with local spin-polarized currents from a scanning tunneling microscope
\cite{Romming-Sci2013}. With a tip of a scanning magnetic force microscope
(MFM), one can cut stripe magnetic domains into skyrmions \cite{Senfu-APL2018}.
It has been suggested that individual skyrmions can also be written
with an MFM tip \cite{AP-2018}. 

Most recently, thermal methods have been explored. 
It was shown that light-induced heat pulses of different
duration and energy can write skyrmions in a magnetic film in
a broad range of temperatures and magnetic fields \cite{Berruto-PRL2018}.
In the last years, skyrmions have been observed at relatively high
temperatures. This poses a question whether temperature alone can
assist the creation of skyrmions. Recently, it has been reported \cite{Zhang2018} that elevated temperature, indeed, facilitates nucleation of skyrmions
from the feromagnetic state and also assists the breaking of magnetic
domains into skyrmions in a Pt/Co/Ta film. In this paper we study this problem theoretically
by Monte-Carlo calculations on large spin lattices. Our model contains
ferromagnetic exchange, Dzyaloshinskii-Moriya interaction (DMI) and
perpendicular magnetic anisotropy (PMA). Some of the calculations
also include random magnetic anisotropy (RMA) to account for disorder
in a real film.

The first comprehensive study of two-dimensional vortex lattices in systems with DMI and uniaxial magnetic anisotropy was done by Bogdanov and Hubert \cite{Bogdanov94} within continuous micromagnetic theory. By considering cylindrically symmetric solutions for individual vortices, they found the dependence of their profile on the strength of the DMI, anisotropy, and the external field. Energies of the periodic arrangement of vortices have been computed and compared with energies of other states to obtain the magnetic phase diagram. More recently the effects of thermal fluctuations and topological charge associated with vortex lattices have been addressed. Yu et al. \cite {Yu2010} used Monte Carlo simulations to explain images of skyrmion phases that they obtained experimentally with the help of the real-space Lorentz TEM of FeCoSi films. By varying temperature and magnetic field they observed transitions between uniformly magnetized states, laminar domains, and skyrmion crystals. A skyrmion tube crystal as a thermodynamically equilibrium phase of a 3D $30 \times 30 \times 30$ anisotropic chiral spin lattice has been found by Monte Carlo method by Buhrandt and Fritz \cite{Buhrandt13}. An important observation made in this paper (that we confirm by Monte Carlo simulations of much larger 2D spin lattices) is that  thermal fluctuations alone can generate a skyrmion state in a narrow region of the field-temperature phase diagram. Stability of individual skyrmions against thermal fluctuations has been studied within the discrete Heisenberg model by Hagemeister et al. \cite{Hagemeister14}. By employing stochastic Landau-Lifshitz equation, R\'ozsa et al.  \cite{Rozsa16} investigated skyrmion lifetime. They also employed Monte Carlo simulations to study creation and annihilation of skyrmions on $128 \times 128$ lattices and computed multi-skyrmion spin configurations, including topological charge. Most recently B$\ddot{\text{o}}$ttcher et al. \cite{Bottcher18} reported parallel-tempering Monte Carlo studies of the temperature dependence of magnetization and topological charge, as well as the field-temperature phase diagram of Pd/Fe/Ir films. They observed formation of both skyrmions and antiskyrmions at high temperatures. 

Our goal is to analyze how thermally assisted skyrmion states depend
on temperature, the external magnetic field, and the PMA  in a basic Heisenberg model with DMI when the field is cycled along the hysteresis loop, without trying to compute the equilibrium phase diagram (that is more of a theoretical abstraction in a magnetic system with hysteresis). It is achieved
by computing the total topological charge along the magnetization
curve on large (up to $1000 \times 1000$) 2D spin lattices and by illustrating skyrmion states with images of the spin
field and images of the topological charge density. We clarify the role of the PMA and compare efficiencies of the static disorder and temperature in creating skyrmions in the magnetization process. We show that fluctuating
spin-field obtained at elevated temperatures can be frozen into a
regular skyrmion lattice by reducing the temperature. Concentration
of skyrmions and their size depend on the external field and on the PMA.
Since the system exhibits profound magnetic hysteresis, with many
local energy minima corresponding to various spin configurations,
it makes sense to compare energies of skyrmion states with energies
of the domain states. In accordance with experiment \cite{Zhang2018} and with previously reported theoretical results,
we find that skyrmion states win in a narrow field range. 

This paper is organized as follows. The model and numerical method
are discussed in Section \ref{Sec_Model}. Spin states emerging along
the magnetization curve at zero temperature are studied in Section
\ref{Sec_States}, where also the field dependence of the topological
charge is computed with and without static disorder in the film. Finite-temperature
effects are studied in Section \ref{Sec_Temp}. In that Section, we
compute the dependence of the hysteresis curves and topological charge
on temperature and PMA, and investigate topology of frozen skyrmion
states. Properties of skyrmion lattices and their energies vs energies
of domain states are computed and compared in Section \ref{Sec_Energy}.
Our conclusions and suggestions for experiment are discussed in Section
\ref{Sec_Discussion}.

\section{The model and numerical method}

\label{Sec_Model} We study the lattice spin model with a Hamiltonian
\begin{eqnarray}
{\cal H} & = & -\frac{J}{2}\sum_{ij}{\bf s}_{i}\cdot{\bf s}_{j}-H\sum_{i}s_{iz}-\frac{D}{2}\sum_{i}s_{iz}^{2}\nonumber \\
 & + & A\sum_{i}[({\bf s}_{i}\times{\bf s}_{i+\hat{x}})\cdot\hat{x}+({\bf s}_{i}\times{\bf s}_{i+\hat{y}})\cdot\hat{y}]\nonumber \\
 & - & \frac{D_{R}}{2}\sum_{i}({\bf n}_{Ri}\cdot{\bf s}_{i})^{2}-\frac{E_{D}}{2}\sum_{ij}\phi_{ij,\alpha\beta}s_{i\alpha}s_{j\beta}\label{Hamiltonian}.
\end{eqnarray}
Here the first term represents the exchange interaction between the
nearest neighbors, with ${\bf s}_{i}$ being the spin at the $i$-th
site of the crystal lattice and $J$ being the exchange constant.
The second term is the Zeeman interaction of the spins with magnetic
field applied in the $z$-direction, perpendicular to the $xy$-plane
of the film. Here $H\equiv g\mu_{B}SB$, $S$ is the value of the
atomic spin, $B$ is the induction of the applied magnetic field, and $g$ is the gyromagnetic factor.
The third term accounts for the PMA of strength $D$. (In our computations the atomic-scale RMA is used, that is, the direction of the RMA axis is assumed random at any lattice site, with its strength being constant. One can also use the model with a correlated RMA that has a stronger effect. The latter allows the use of a smaller anisotropy constant to achieve a similar effect.) The fourth term
describes the Bloch-type DMI \cite{Leonov-NJP2016} of strength $A$.
For the N\'eel-type DMI it should be replaced with $A\sum_{i}[({\bf s}_{i}\times{\bf s}_{i+\hat{x}})\cdot\hat{y}-({\bf s}_{i}\times{\bf s}_{i+\hat{y}})\cdot\hat{x}]$.
The fifth term incorporates the effect of the RMA of strength $D_{R}$,
with ${\bf n}_{Ri}$ describing a randomly chosen orientation of the
anisotropy axis at the $i$-th lattice site. The last term is the
DDI with
\begin{equation}
\phi_{ij,\alpha\beta}\equiv a^{3}r_{ij}^{-5}\left(3r_{ij,\alpha}r_{ij,\beta}-\delta_{\alpha\beta}r_{ij}^{2}\right).
\end{equation}
Here $r_{ij}$ is the distance between the $i$-th and the $j$-th
site, $a$ is the lattice spacing, $\alpha,\beta=x,y,z$ denotes spin
components, and $E_{D}=\mu_{0}M_{0}^{2}a^{3}/(4\pi)$ is the strength
of the DDI (with $\mu_{0}$ being the magnetic permeability of vacuum
and $M_{0}$ being the saturation magnetization, $M_{0}=g\mu_{B}S/a^{3}$
for our lattice model). In all our computations, we have found that
the effect of the DMI on spin configurations dominates over the competing
effect of a much weaker DDI.

To describe non-uniform spin states, one needs to consider a system
with a large number of spins. In the lattice model, this leads to
long computation times. For many materials the problem is further exacerbated by the weakness
such interactions as DMI, Zeeman, PMA, RMA, and DDI in
comparison with the ferromagnetic exchange. It makes the system
magnetically soft, further slowing down the convergence towards the
energy minimum in the computation routine. In such materials the emerging
spin structures are large compared to the atomic spacing $a$. When interactions are small, to
speed up computation, one can rescale the 2D
problem to the same problem with a bigger lattice constant $b>a$
and new parameters $J'=J$,\, $A'=(b/a)A$,\, $H'=(b/a)^{2}H$,\,
$D'=(b/a)^{2}D$. The rescaled problem has a smaller number of mesh
points $N_{\alpha}'=(a/b)N_{\alpha}$ and smaller mismatch between
$J$ and other interaction constants. The results of the computation
can be easily rescaled back to the original model. 

In what follows we use $A/J=0.2$, $D/J=0$ or $0.01$ or $0.02$, and $D_{R}/J=0$
or $0.3$, and set $J=1$, $s=1$. Calculations have been performed
on 2D lattices of size up to $300\times300$.

Among other features we are interested in the topological charge of
the emerging spin structures. In a continuous spin-field model it
is given by \cite{BelPolJETP75}
\begin{equation}
Q=\int dxdy\,q(x,y),\,\quad q(x,y)=\frac{1}{4\pi}\:{\bf s}\cdot\frac{\partial{\bf s}}{\partial x}\times\frac{\partial{\bf s}}{\partial y}.\label{Q}
\end{equation}
The number $Q$ takes discrete values $Q=0,\pm1,\pm2,...$. The function
$q(x,y)$ can be interpreted as the density of the topological charge.
It provides a better visualization of the location of skyrmions and
antiskyrmions when they are strongly deformed from the minimum-energy
Belavin-Polyakov (BP) shape by disorder and competing interactions and cannot be easily identified by looking at the spin field
\cite{EC-DG-PRL2018}. In numerical work we use a lattice-discretized version
of Eq.\ (\ref{Q}).

\begin{figure}
\begin{centering}
\includegraphics[width=9cm]{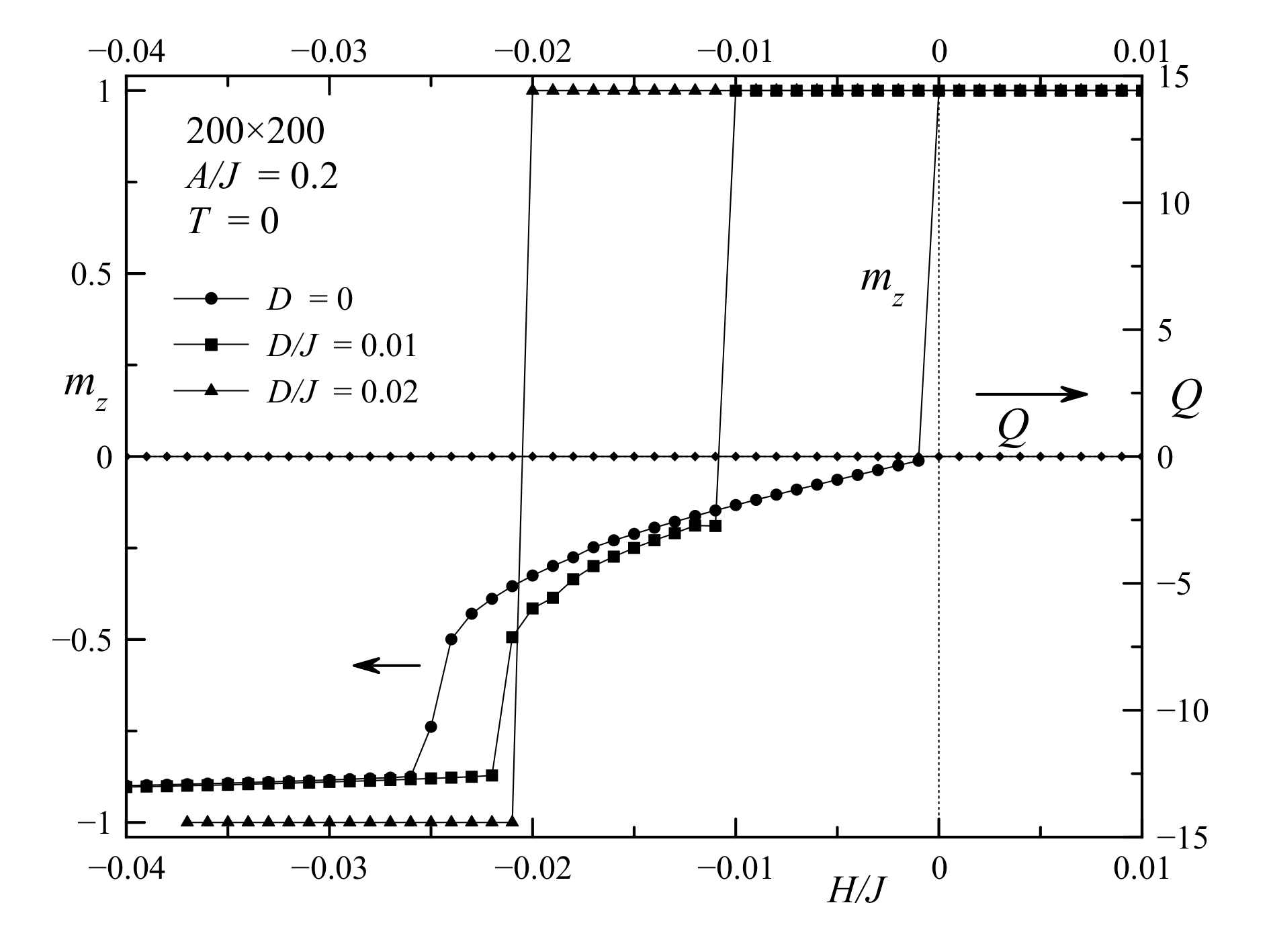}
\par\end{centering}
\caption{Magnetization curves and topological charge at $T=0$ in the model
with DMI and PMA.}
\label{MagT=00003D0}
\end{figure}
At zero temperature, we compute minimum-energy configurations of spins.
Our numerical method combines sequential rotations of spins ${\bf s}_{i}$
towards the direction of the local effective field, ${\bf H}_{{\rm eff},i}=-\partial{\cal H}/\partial{\bf s}_{i}$,
with the probability $\alpha$, and the energy-conserving spin flips
(so-called \textit{overrelaxation}), ${\bf s}_{i}\to2({\bf s}_{i}\cdot{\bf H}_{{\rm eff},i}){\bf H}_{{\rm eff},i}/H_{{\rm eff},i}^{2}-{\bf s}_{i}$,
with the probability $1-\alpha$. The parameter $\alpha$ plays the
role of the effective relaxation constant. We mainly use the value
$\alpha=0.03$ that provides the overall fastest convergence. At nonzero
temperatures, we replace the rotation of spins towards the direction
of the effective field by Monte Carlo updates, keeping the over-relaxation
dominant via the same small value of $\alpha$. We change the field in small steps and allow the system to reach (metastable) equilibrium at each step. This is equivalent to a slow field sweep in experiment. 
Due to the large system size, thermodynamic fluctuations and the statistical
scatter are small, making the predicted features reliable with a large
degree of confidence.

\section{Spin states at zero temperature}

\label{Sec_States}

\begin{figure}
\begin{centering}
\includegraphics[width=9cm]{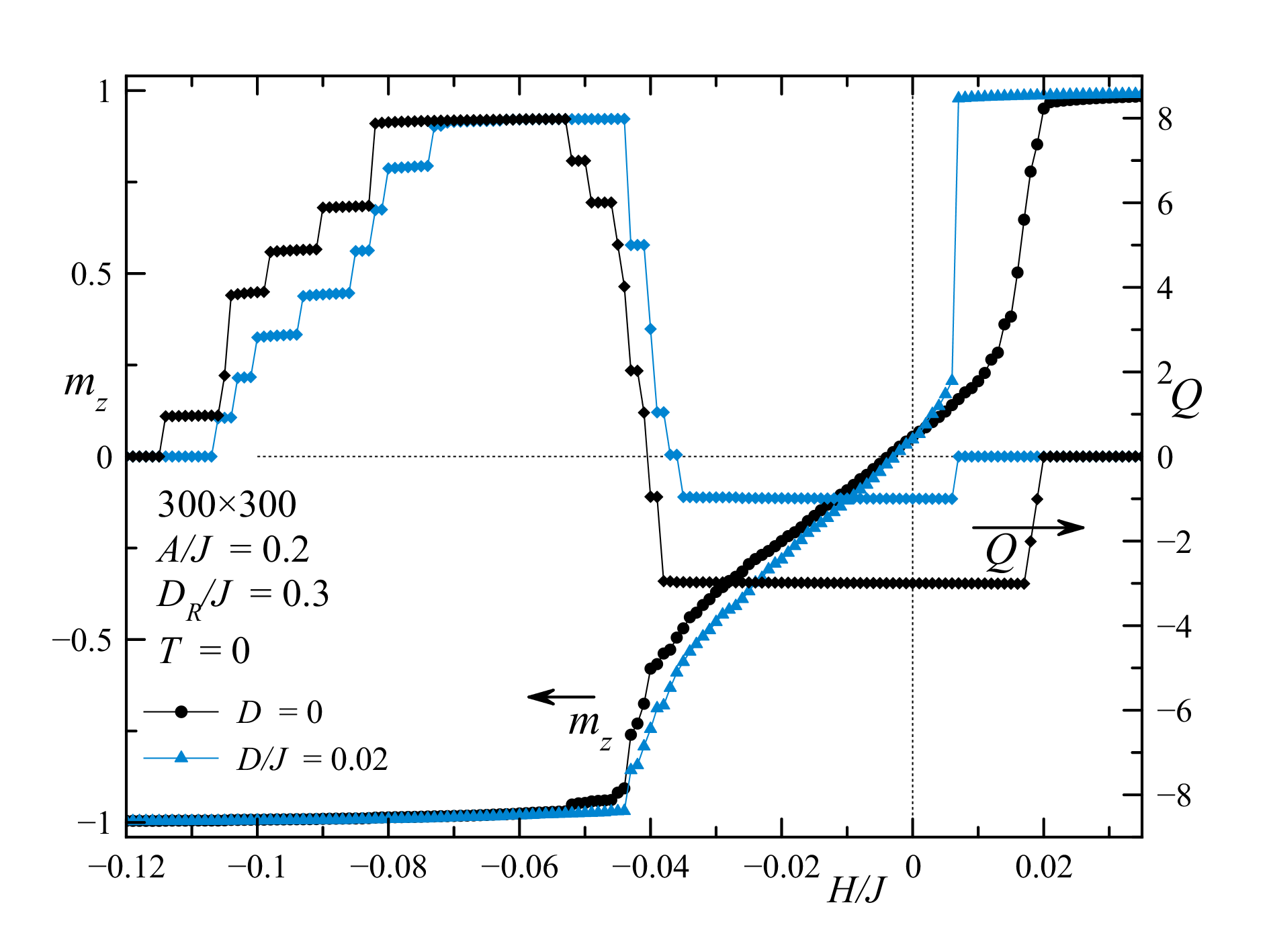}
\includegraphics[width=9cm]{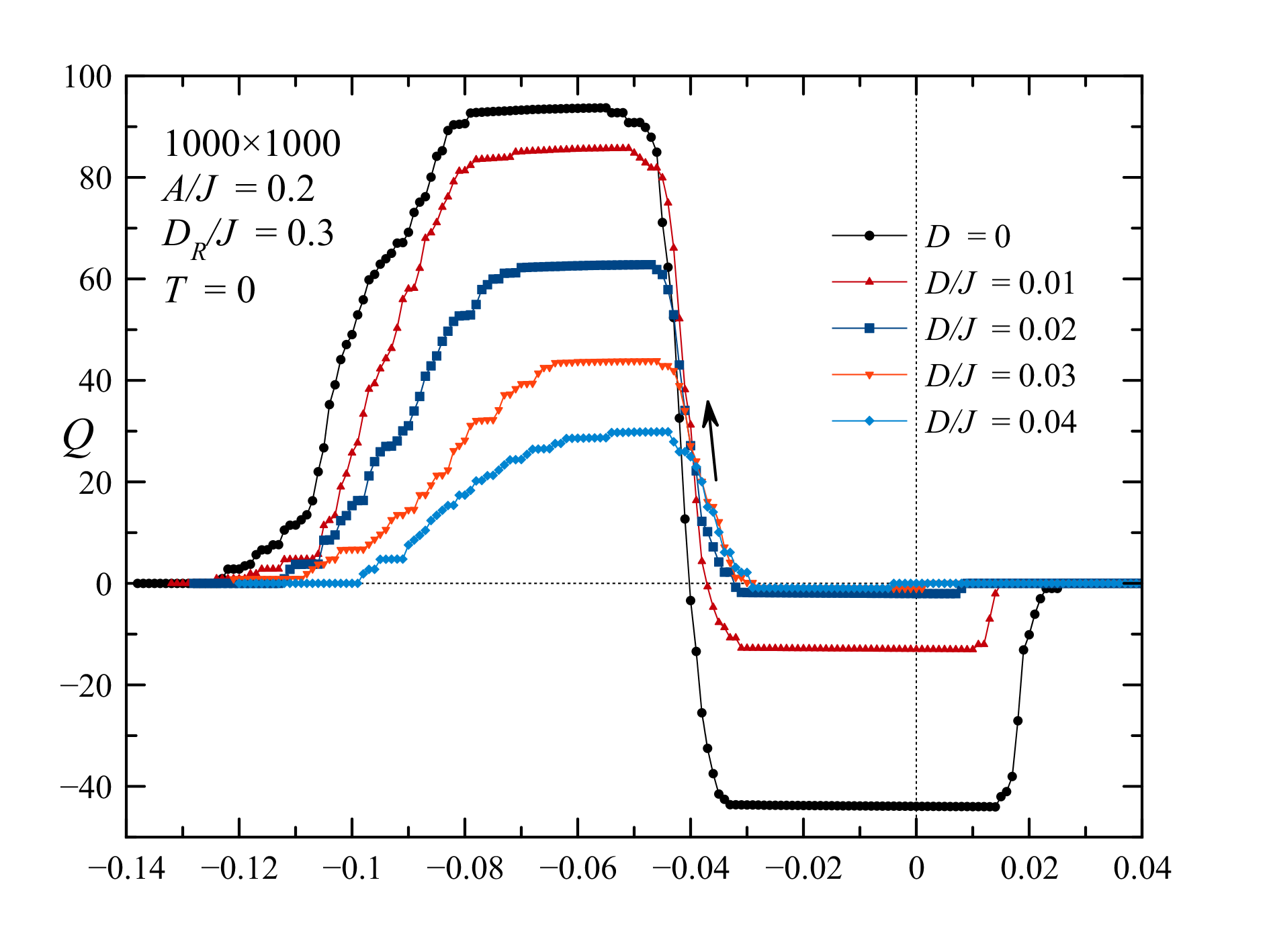}
\par\end{centering}
\caption{Magnetization curves and topological charge at $T=0$ in the model
with DMI, PMA, and RMA. Upper panel: $300 \times 300$ lattice. Lower panel: Topological charge in a $1000 \times 1000$ lattice with many more skyrmions. }
\label{RMA-T=00003D0}
\end{figure}
Fig.\ \ref{MagT=00003D0} shows the numerical results for the reduced
magnetization per spin $m_z = \langle s_{z} \rangle$ (left axis), with the average taken over all lattice sites, and the total topological
charge $Q$ (right axis) at $T=0$. As $H$ decreases from saturation,
the uniform state with $m_{z}=1$ becomes unstable and the system
jumps into the state with laminar domains created by the DMI. As the
field continues to change, this domain structure becomes gradually
suppressed, leading to the uniformly magnetized state with $m_{z}=-1$
at the field saturating the system in the negative $z$-direction.
In the case of zero anisotropy, $D=0$, the instability of the uniformly
magnetized state occurs exactly at $H=0$. Finite PMA makes the uniform
state more stable, so that the instability occurs at $H=-D$ when
the applied field compensates the effective anisotropy field. One
can see that $Q=0$ at $T=0$ everywhere along the magnetization curve.
This is a consequence of the conservation of the topological charge in
the absence of thermal or static randomness.

\begin{figure}
\begin{centering}
\includegraphics[width=8cm]{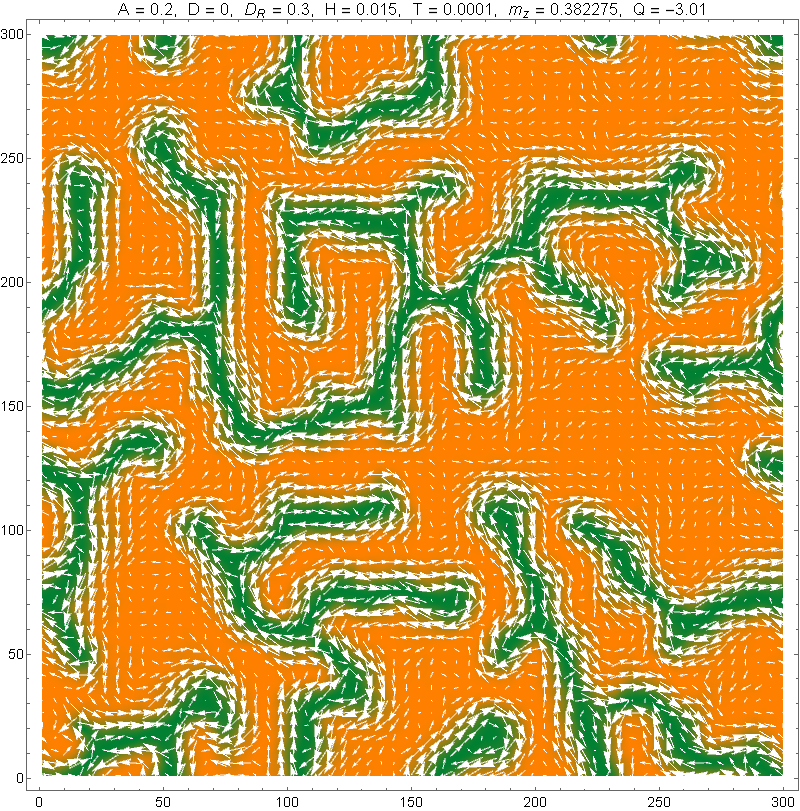}
\par\end{centering}
\caption{Proliferation of worm-like domains at $T=0$ in the model with DMI and RMA at $H > 0$. Orange/green color indicates positive/negative
$z$-component of the spin field. The in-plane spin components are shown as white arrows.}
\label{labyrinth}
\end{figure}

\begin{figure}
\begin{centering}
\includegraphics[width=8cm]{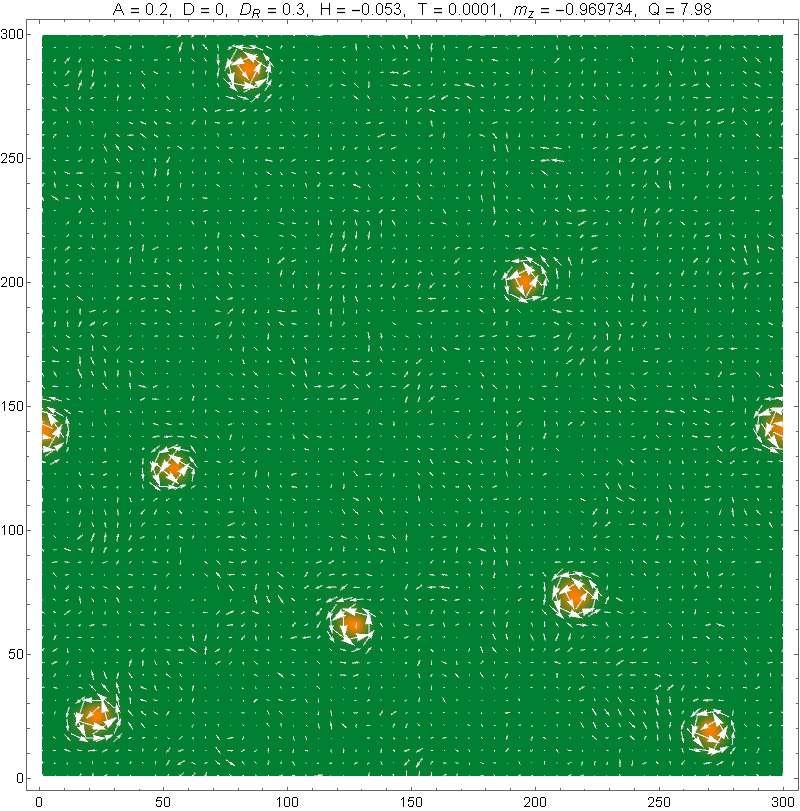}
\par\end{centering}
\caption{Small number of skyrmions resulting from the breaking of labyrinth
domains by the magnetic field, $H < 0$,  in the presence of RMA at $T=0$.}
\label{few-skyrmions}
\end{figure}
The effect of static randomness at $T=0$, generated by the RMA, is
illustrated in Fig. \ref{RMA-T=00003D0}. One can see that in the
absence of the PMA, the RMA is triggering the instability of the quasi-uniformly
magnetized state with $m_{z}\approx1$ already at $H>0$. The destruction
of the uniformly magnetized state leads to the proliferation of worm-like domains, see  Fig.\ \ref{labyrinth}, with a
non-zero though small topological charge per unit area. According to Fig.\ \ref{RMA-T=00003D0} a non-zero PMA with
$D/J=0.02$ increases the stability range of the uniformly magnetized
state and reduces $Q$ of the domain state even further. The scaling of the effect with the area of the film is illustrated by the lower panel of Fig.\ \ref{RMA-T=00003D0} that shows a significantly greater $Q$ and a smoother dependence on the field due to many more skyrmions generated in a larger $1000 \times 1000$ spin lattice.

\begin{figure}
\begin{centering}
\includegraphics[width=9cm]{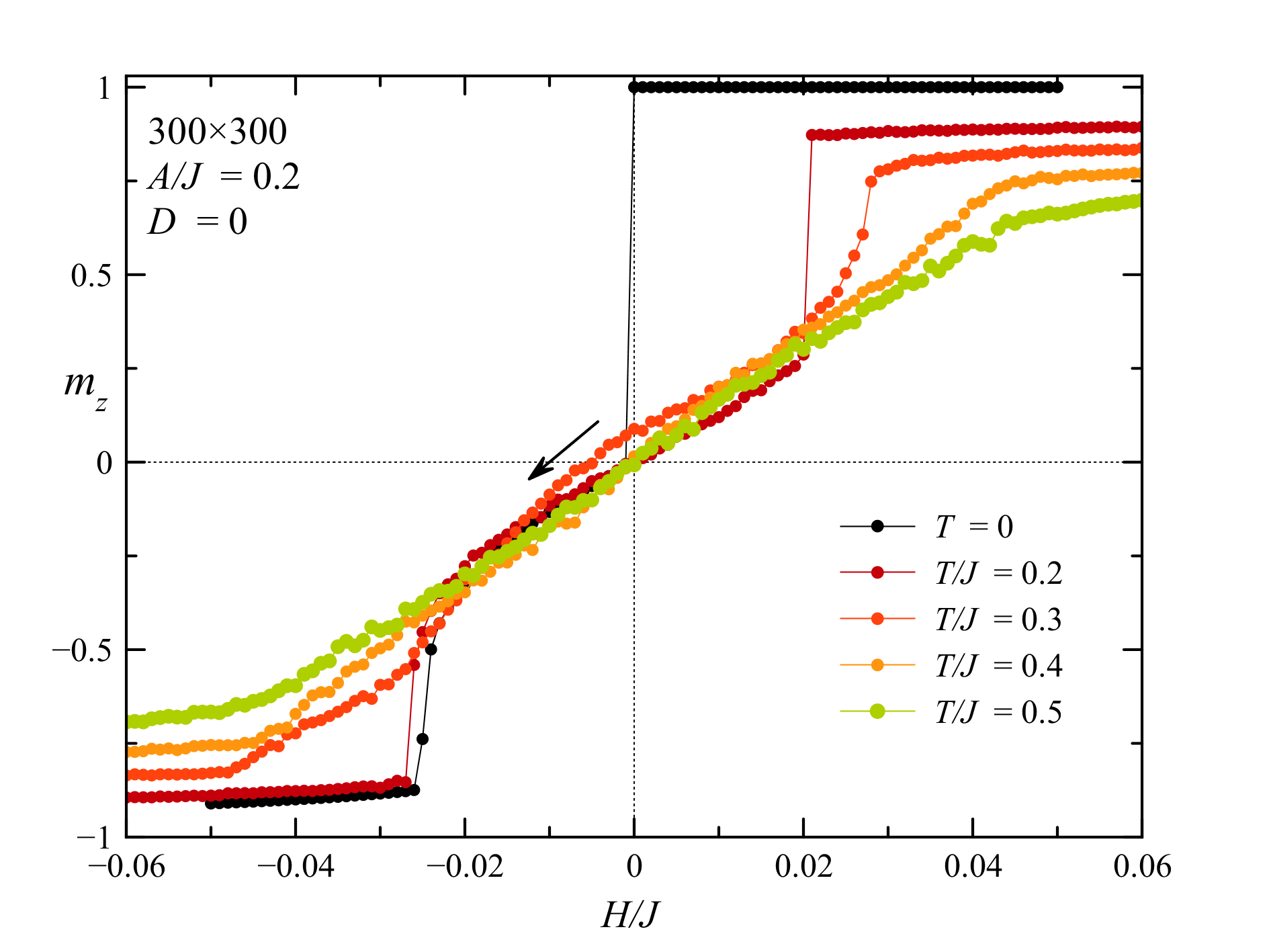} \includegraphics[width=9cm]{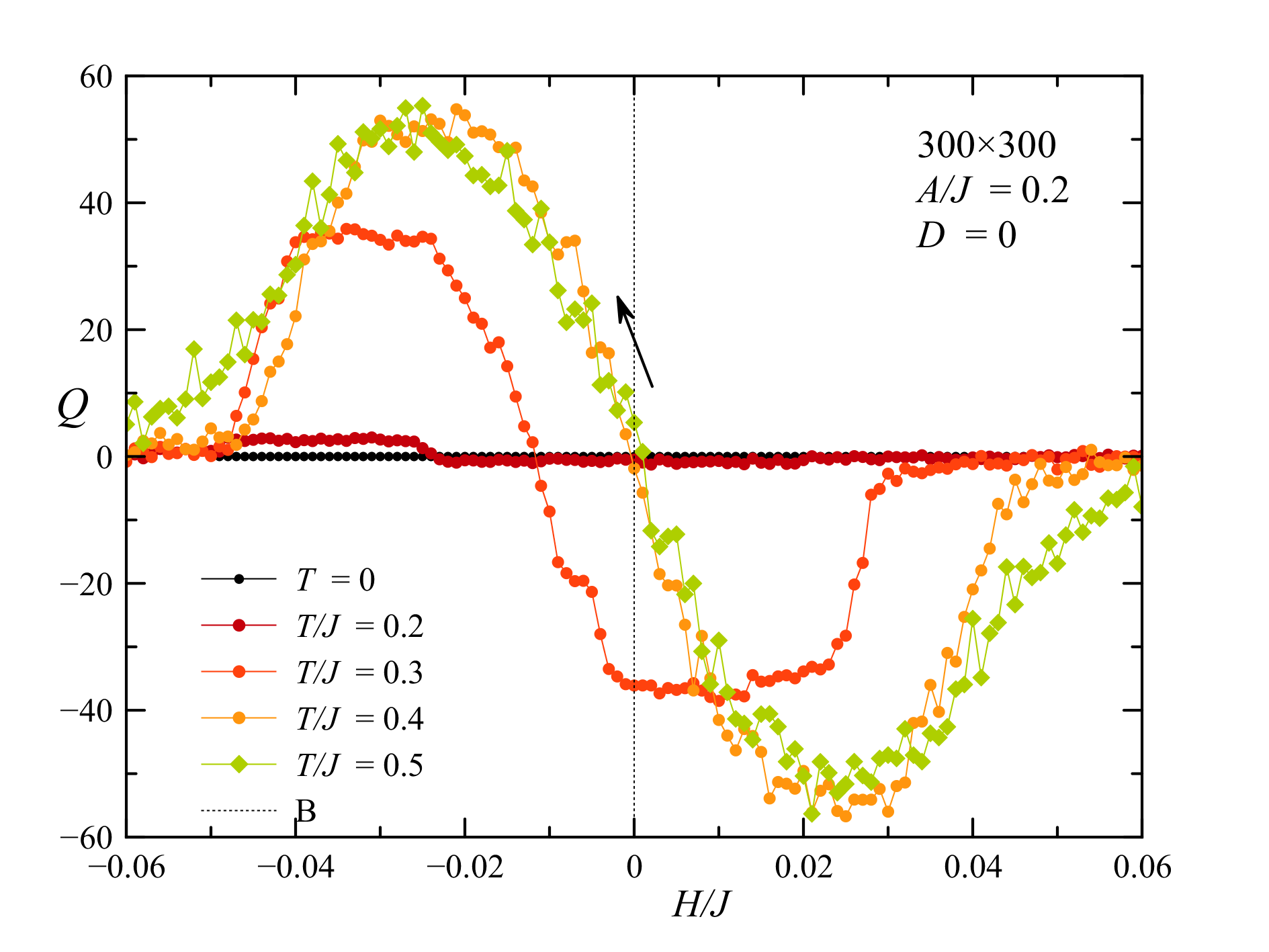}
\par\end{centering}
\caption{Magnetization curves (upper panel) and topological charge (lower panel)
at finite temperature in the model with DMI at $D=0$.}
\label{finiteT-D=00003D0}
\end{figure}
As the magnetic field switches towards the negative direction, while
changing continuously along the magnetization curve (see Fig.\ \ref{RMA-T=00003D0}),
some domains break into skyrmions, creating a few skyrmions shown
in Fig.\ \ref{few-skyrmions}. While skyrmions do not appear spontaneously at $T = 0$ in the absence of RMA, the maximum number of skyrmions emerging
in the magnetization process from labyrinth domains in the presence of RMA is still small at $T=0$. As we shall see below, temperature has a much more profound effect on the emergence of skyrmions.  

\section{Spin states at finite temperature}

\label{Sec_Temp}

\begin{figure}
\begin{centering}
\includegraphics[width=8cm]{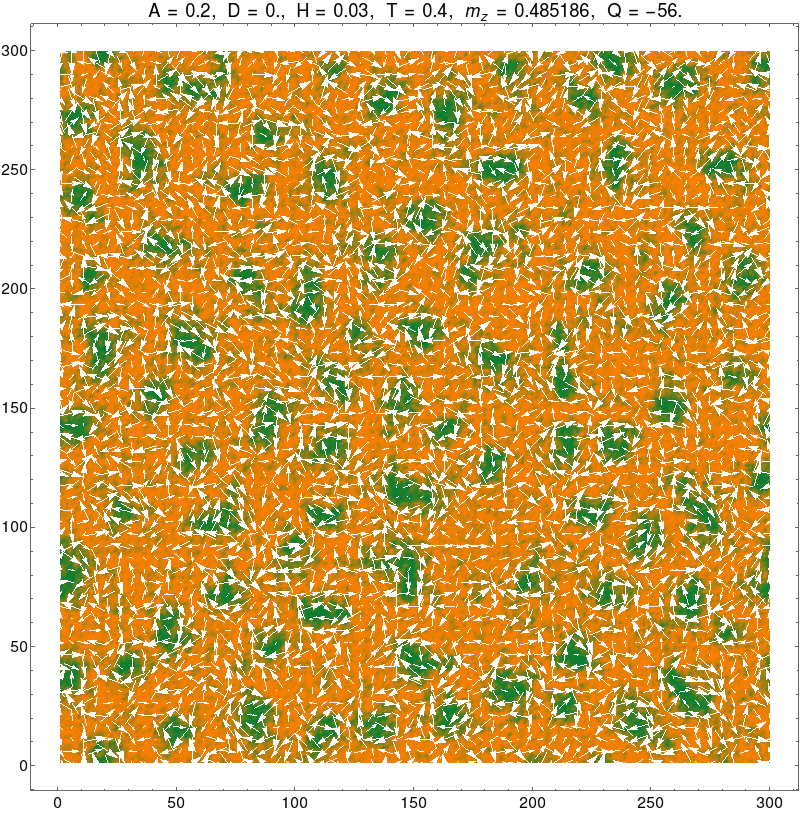}
\par\end{centering}
\caption{Thermally nucleated skyrmions at $T/J=0.4$, $D=0$, and $H/J=0.03$.}
\label{finiteT-H=00003D0.03}
\end{figure}
We will now turn to the creation of skyrmions at finite temperature.
The effect of temperature is two-fold. Firstly, thermal fluctuations
violate conservation of the topological charge even in the pure exchange
model. Secondly, spin fluctuations reduce the effective PMA that tends
to suppress skyrmions. As we shall see, they first emerge via thermal nucleation
from the uniformly magnetized state and then when the destruction of ferromagnetic
domains by the magnetic field is assisted by temperature.
\begin{figure}
\begin{centering}
\includegraphics[width=8cm]{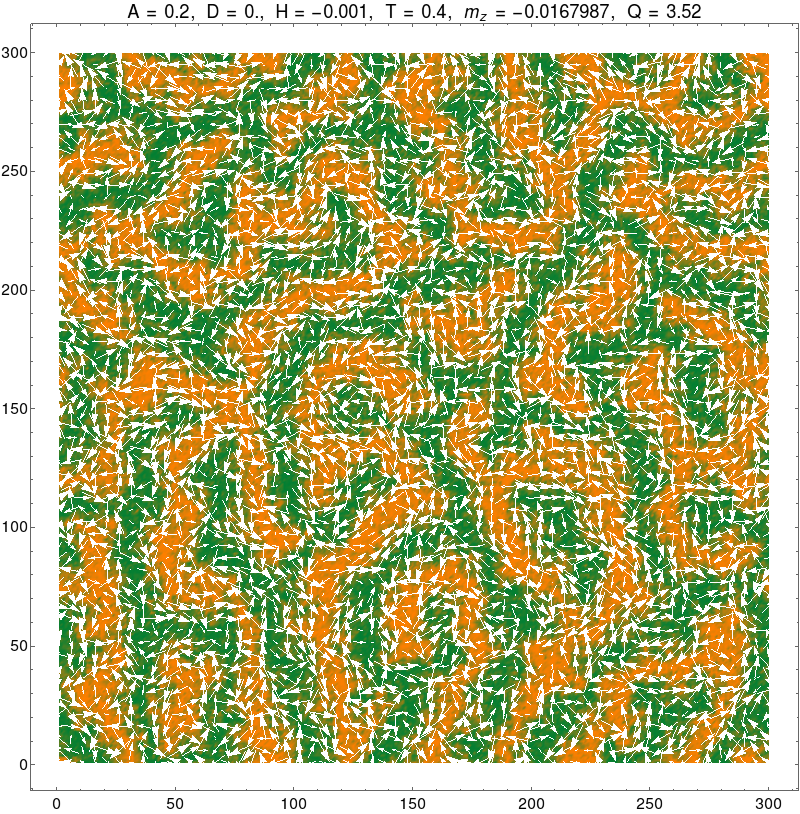}
\par\end{centering}
\caption{Labyrinth domains from merging skyrmions at $T/J=0.4$, $D=0$, and
$H/J=-0.001$.}
\label{finiteT-H=00003D-0.001}
\end{figure}

\begin{figure}
\begin{centering}
\includegraphics[width=8cm]{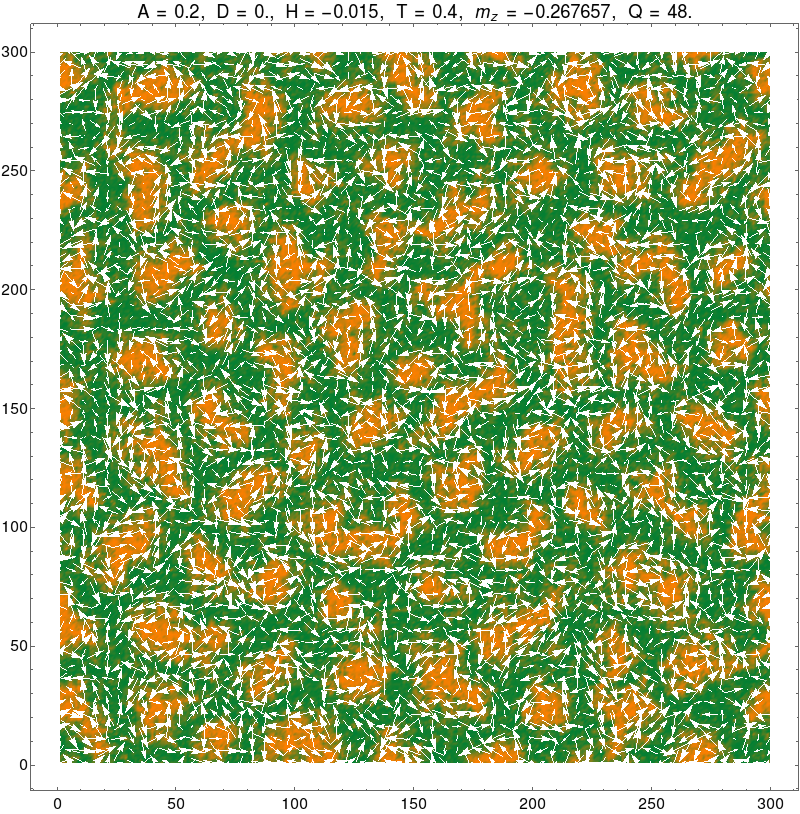}
\par\end{centering}
\caption{Thermally assisted formation of skyrmions from labyrinth domains broken
by the field at $T/J=0.4$, $D=0$, and $H/J=-0.015$.}
\label{finiteT-H=00003D-0.015}
\end{figure}
Fig. \ref{finiteT-D=00003D0} shows the dependence of $m_{z}(H)$
and $Q(H)$ in the model with DMI and no PMA at different temperatures.
At $T/J=0.5$ and $0.4$ the system is approximately in thermal equilibrium
at any $H$, so that the magnetization curve is reversible. At lower
temperatures, the equilibrium cannot be achieved within a reasonable
time of Monte Carlo simulations and the reversibility is lost, which
reflects the physics of the magnetic hysteresis. The value of $T/J=0.3$
is an intermediate one for which a partial thermal equilibrium is
achieved. One can see that the number of created skyrmions decreases
at low temperatures, so that the results presented in Fig.\ \ref{MagT=00003D0}
are recovered.

\begin{figure}
\begin{centering}
\includegraphics[width=9cm]{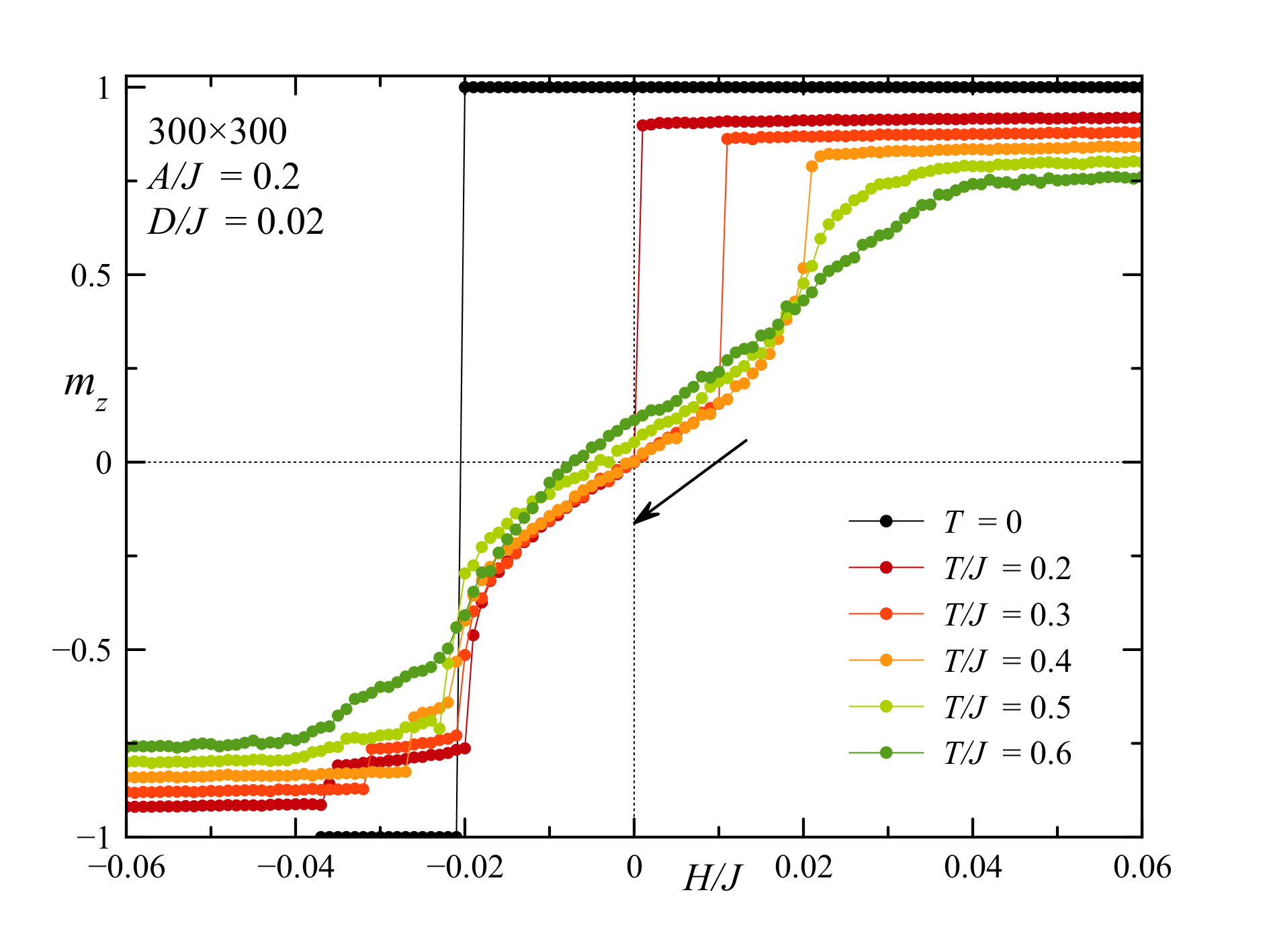} \includegraphics[width=9cm]{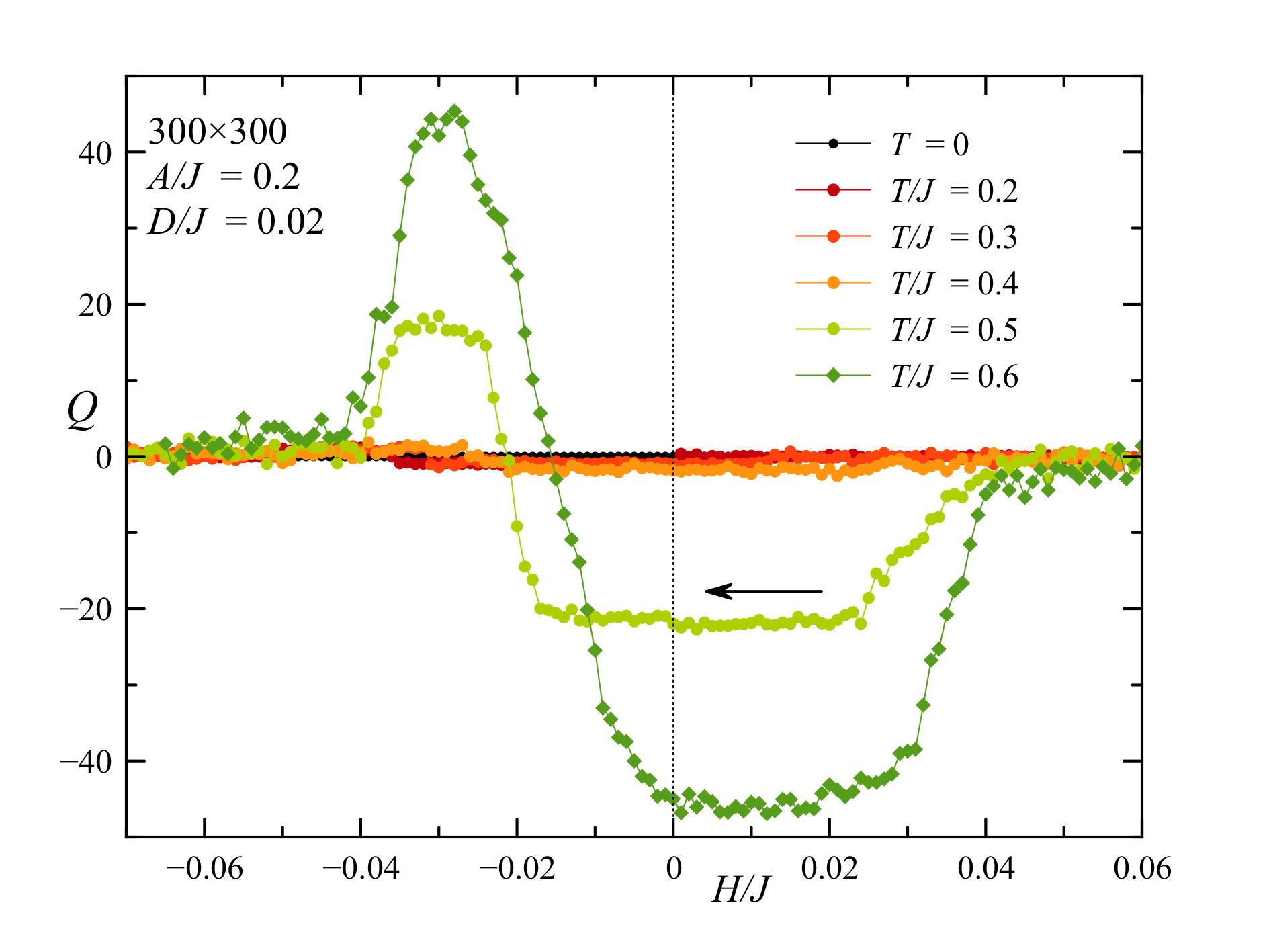}
\par\end{centering}
\caption{Magnetization curves (upper panel) and topological charge (lower panel)
at finite temperature in the model with DMI and PMA.}
\label{finiteTandD}
\end{figure}
Finite temperatures cause nucleation of skyrmions at $H>0$ with a rather
large topological charge $Q$ of the entire film, see Fig. \ref{finiteT-H=00003D0.03}.
As the magnetic moments of these skyrmions point down against the
spin-up background, $Q$ is negative. Decreasing the field towards
$H\approx0$ increases skyrmion concentration until skyrmions begin
to merge into magnetic domains, see Fig.\ \ref{finiteT-H=00003D-0.001},
and the topological charge of the film drops. As the field increases
in the negative direction, the domains begin to break back into skyrmions,
making $Q$ positive and large again, see Fig.\ \ref{finiteT-H=00003D-0.015}.
Skyrmions collapse as the magnetization of the film approaches saturation. Note that at elevated temperatures spin configurations fluctuate. The pictures above show snapshots of such configurations. 

Fig. \ref{finiteTandD} shows the results of similar computations
with a finite PMA of strength $D/J=0.02$. The latter strongly suppresses
skyrmions, their number becoming significant only at elevated temperatures
such as $T/J=0.5$, when thermodynamic equilibrium in Monte Carlo
simulations is still incomplete.
\begin{figure}
\begin{centering}
\includegraphics[width=8cm]{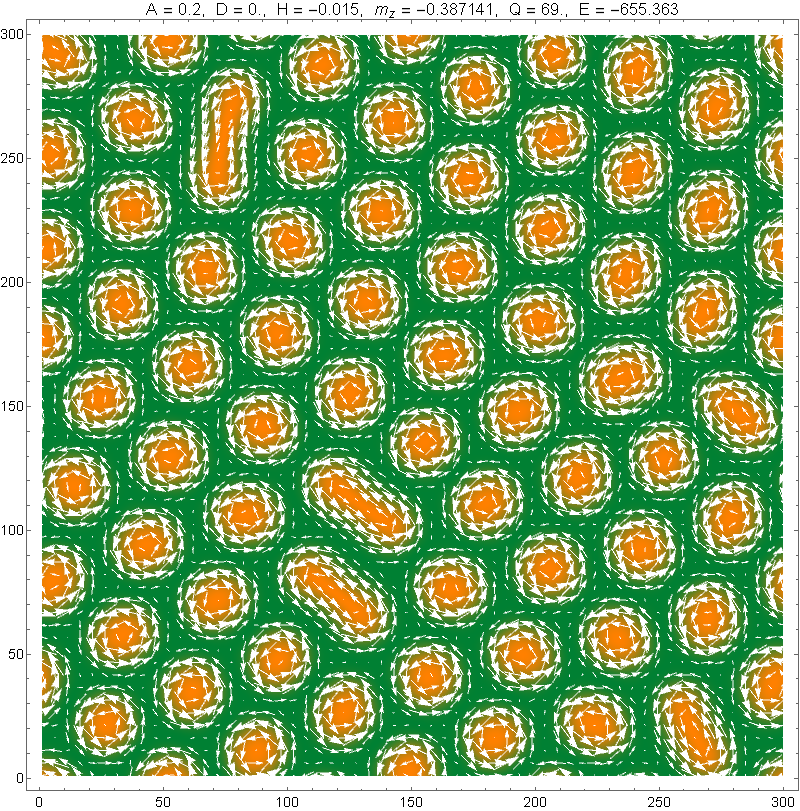}
\par\end{centering}
\caption{Skyrmion lattice obtained by freezing down to $T=0$ the thermally
assisted skyrmion state shown in Fig.\ \ref{finiteT-H=00003D-0.015}.}
\label{freezingSmall-H}
\end{figure}

\begin{figure}
\begin{centering}
\includegraphics[width=8cm]{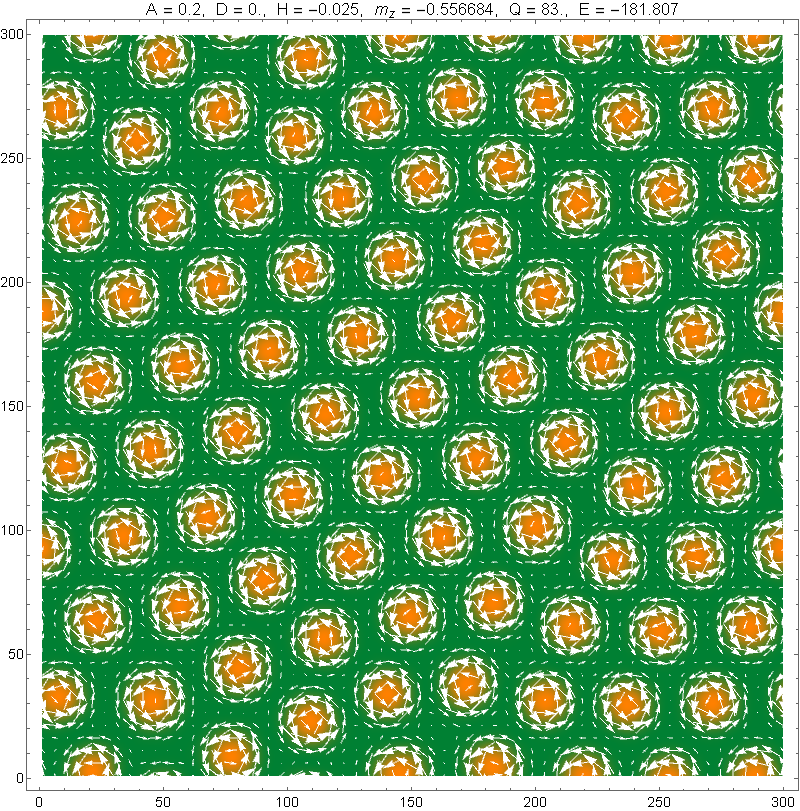}
\par\end{centering}
\caption{Skyrmion lattice obtained at $H/J=-0.025$ by freezing down to $T=0$
the random-spin state at $T=\infty$. }
\label{freezingLarge-H}
\end{figure}
We also have performed numerical experiments that simulate freezing
of the fluctuating skyrmion states obtained at elevated temperatures
down to $T=0$, which is similar to the protocol of some real experiments with skyrmions \cite{Milde2013,Oike2015}.  An example of the freezing of the skyrmion state generated
at $H/J=-0.015$ and $T/J=0.4$ (see Fig. \ref{finiteT-H=00003D-0.015})
is shown in Fig.\ \ref{freezingSmall-H}. In this example, the freezing
leads to an imperfect skyrmion lattice. Similar numerical experiment
at $H/J=-0.025$ and $T=0$, obtained from the random-spin state at
$T=\infty$ as the initial condition for the relaxation, leads to
a more regular skyrmion lattice with a larger number of smaller skyrmions,
see Fig. \ref{freezingLarge-H}.

\section{Skyrmion lattices vs labyrinth domains}

\label{Sec_Energy}

\begin{figure}
\begin{centering}
\includegraphics[width=8cm]{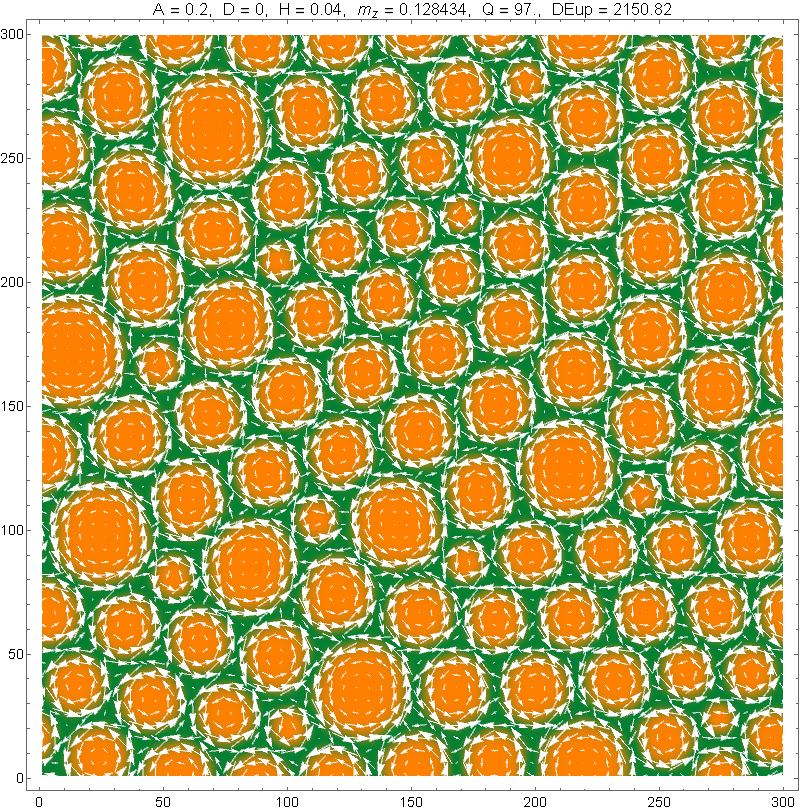}
\par\end{centering}
\caption{Metastable skyrmion state that evolved on increasing the field from the skyrmion
lattice at $H/J=-0.025$ shown in Fig. \ref{freezingLarge-H}. The
skyrmion structure is shown at $H/J=0.04$ that is close to the critical
field at which the walls between the skyrmion bubbles collapse.}
\label{PackedSkyrmions}
\end{figure}
In a different numerical experiment, we investigated skyrmion states
at $T=0$ evolving on changing the applied field $H$ in both directions
in small steps starting from the state shown in Fig. \ref{freezingLarge-H}.
Initially, 90-100 skyrmions
emerged in the $300\times300$ lattice at $H/J=-0.025$. Decreasing
the field from the initial value, makes skyrmions smaller and eventually
leads to their collapse. Increasing the field makes skyrmions larger.
When they come in contact with each other, the regular skyrmion lattice
transforms into densely packed skyrmions of various size shown in
Fig.\ \ref{PackedSkyrmions}. On further increase of the field the walls separating skyrmion bubbles become
thinner. As the field continues to grow, skyrmion walls collapse in a single large jump leading to a quasi-uniform magnetization.

\begin{figure}
\begin{centering}
\includegraphics[width=8cm]{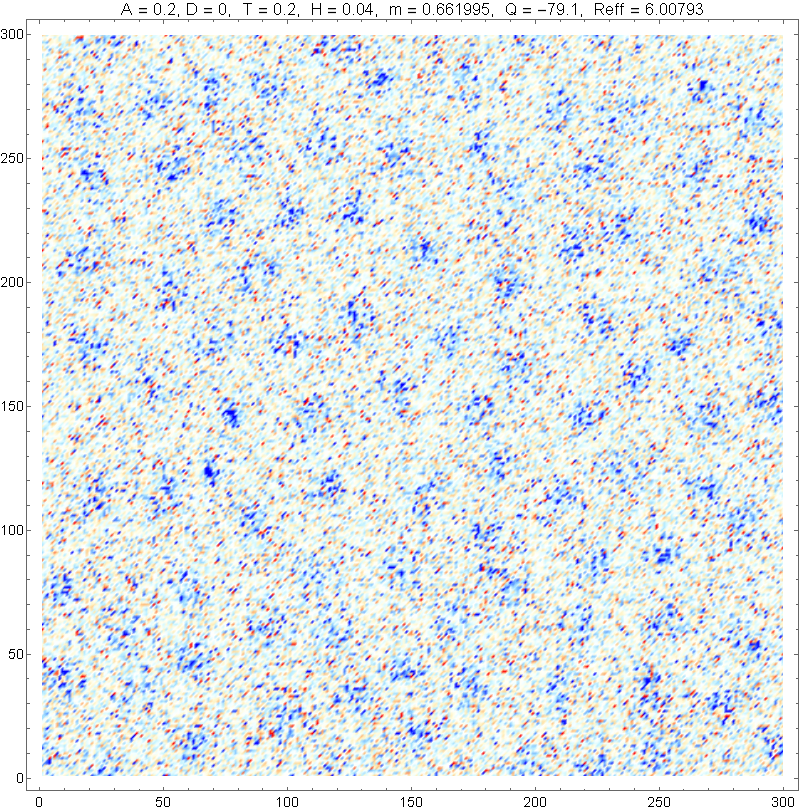} \includegraphics[width=8cm]{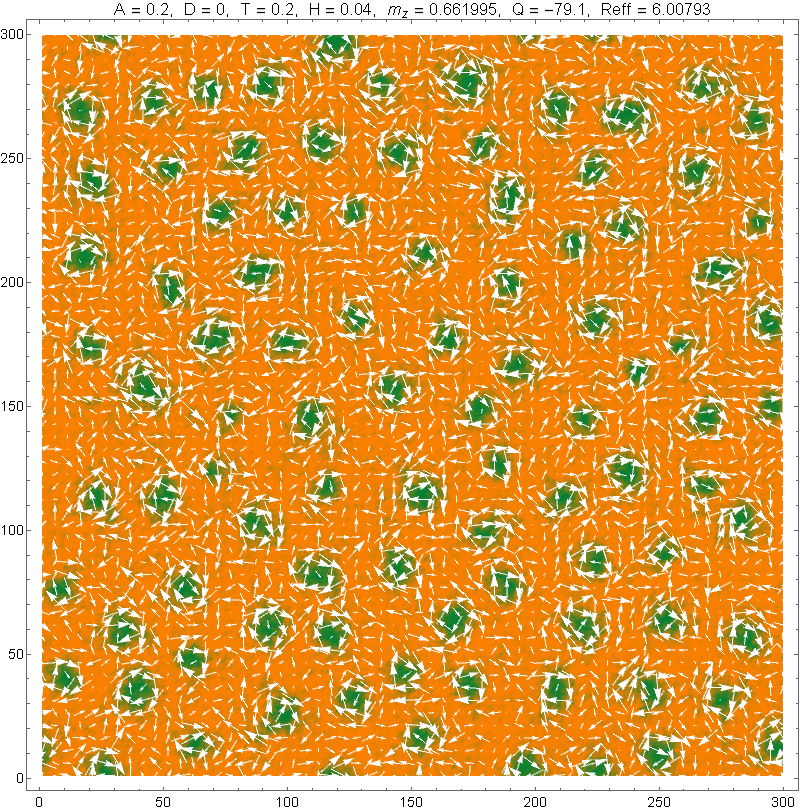}
\par\end{centering}
\caption{Topological charge density (TCD) (upper panel) for the skyrmion lattice
shown in the lower panel. The color code in the TCD plot is red for
positive TCD and blue for negative TCD. }
\label{TCD}
\end{figure}
In a systems with a strong DMI, the number of skyrmions at $T=0$
equals $Q$ and can be counted easily. The effective radius of
the skyrmion bubble that does not have a BP shape can be estimated
using the formula for a thin-wall bubble,
\begin{equation}
\pi QR_{\mathrm{eff}}^{2}=\frac{1}{2}\int\int dxdy\left[1\pm s_{z}(x,y)\right],\label{R}
\end{equation}
if $s_{z}=\mp1$ in the background. To make this formula more robust
for elevated temperatures, when $s_{z}$ in the background deviates
from $\pm1$ due to thermal fluctations, one can replace $s_{z}(x,y)$ in Eq.\ (\ref{R}) with $\pm 1$ depending on its sign.
However, at elevated temperatures, while $Q$ remains a well-defined reproducible quantity, there is a significant washing-out of the TCD, as Eq. (\ref{Q}) contains
two derivatives and the spin field changes at the lattice scale. This
is confirmed by looking at the density of the topological charge plotted
in the upper panel of Fig.\ \ref{TCD} for the skyrmion lattice shown
in the lower panel of that figure. At higher temperatures, the TCD
washes out completely due to strong spin fluctuations at the atomic-scale. Still, the integral value of $Q$ turns out
to be robust and consistent with the number of skyrmions one can count in the image of spin field similar to the one shown 
in the lower panel of Fig.\ \ref{TCD}.

\begin{figure}
\begin{centering}
\includegraphics[width=9cm]{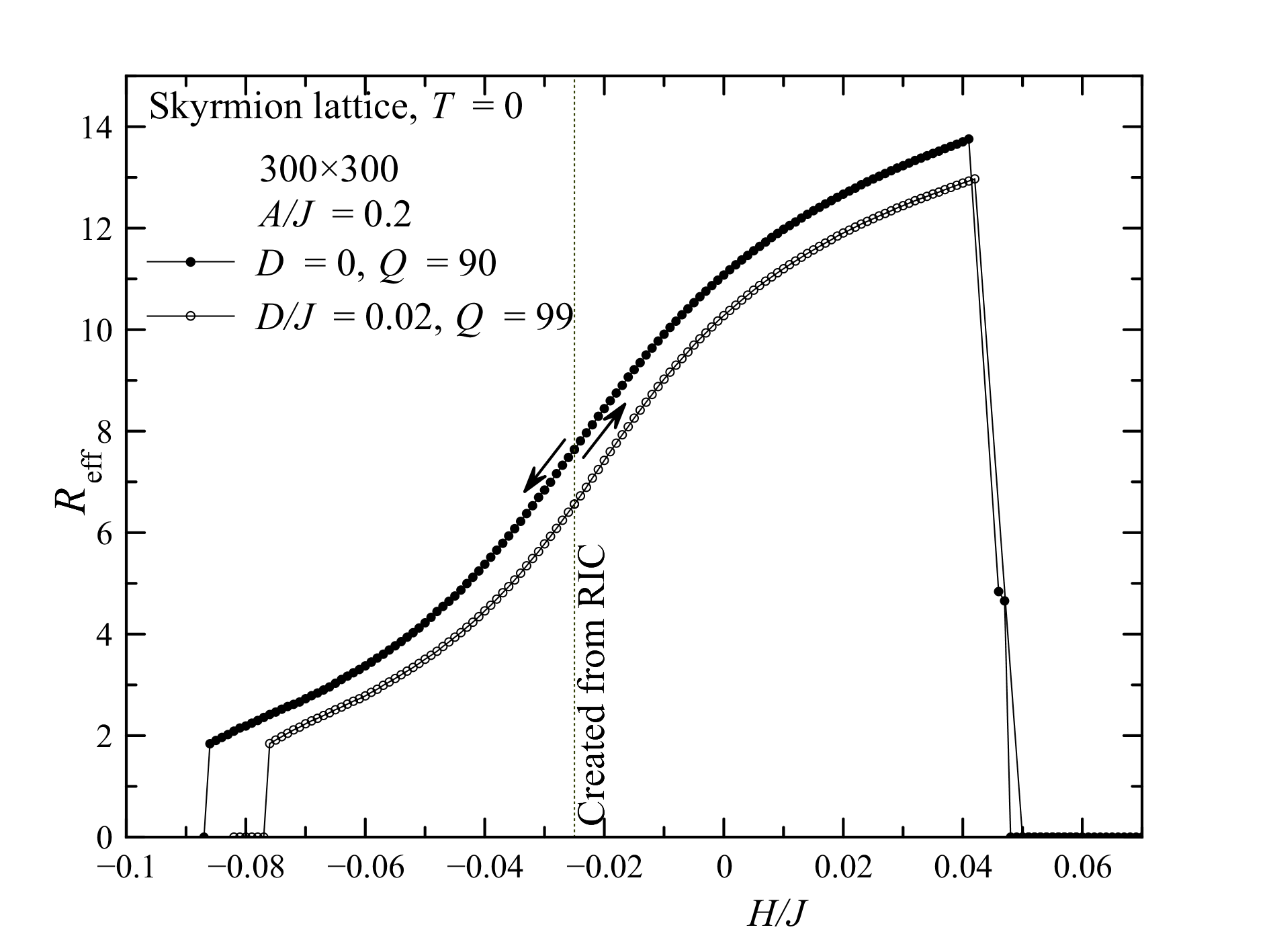}
\par\end{centering}
\caption{Field dependence of the effective radius of skyrmions in the skyrmion
lattice created from the RIC at $T=0$ with $D/J=0$ and $0.02$ on
increasing and decreasing the field from the initial value $H/J=-0.025$.}
\label{Reff}
\end{figure}
The dependence of $R_{\mathrm{eff}}$ on $H$ for $D/J=0.02$ and
0 is shown in Fig.\ \ref{Reff}. For the model with anisotropy $R_{\mathrm{eff}}$
is smaller and the collapse on the left side occurs earlier. This
can be understood in simple terms within the continuous spin-field
model. If one neglects the dependence of the exchange energy on skyrmion
size and neglects interaction between skyrmions, the DMI energy of
the skyrmion scales as $-AR_{{\rm eff}}$ while the PMA+ Zeeman energy
scales as $(D+\left|H\right|)R_{{\rm eff}}^{2}$. (This scaling of the PMA assumes $R_{\rm eff} < \delta$, where $\delta = \sqrt{J/D}$ is the domain wall width.) The energy minimum is achieved at $R_{{\rm eff}}\propto A/(D+\left|H\right|)$, which
agrees qualitatively with Fig.\ \ref{Reff} as long as $H<0$ and
is sufficiently strong. At $D=0$ and $H\rightarrow0$, the formula for
$R_{{\rm eff}}$ above diverges. Before it happens, skyrmions evolve into labyrinth
domains. The skyrmion lattice at $H>0$ is stable due to the interaction between the skyrmions.
On increasing the field the skyrmions grow and become densely packed, see Fig.\ \ref{PackedSkyrmions}. They burst at the well defined field,  leading to the uniform magnetization. This corresponds to the right-hand end of the $R_{\rm eff}(H)$ curve in Fig.\ \ref{Reff}.

\begin{figure}
\begin{centering}
\includegraphics[width=8cm]{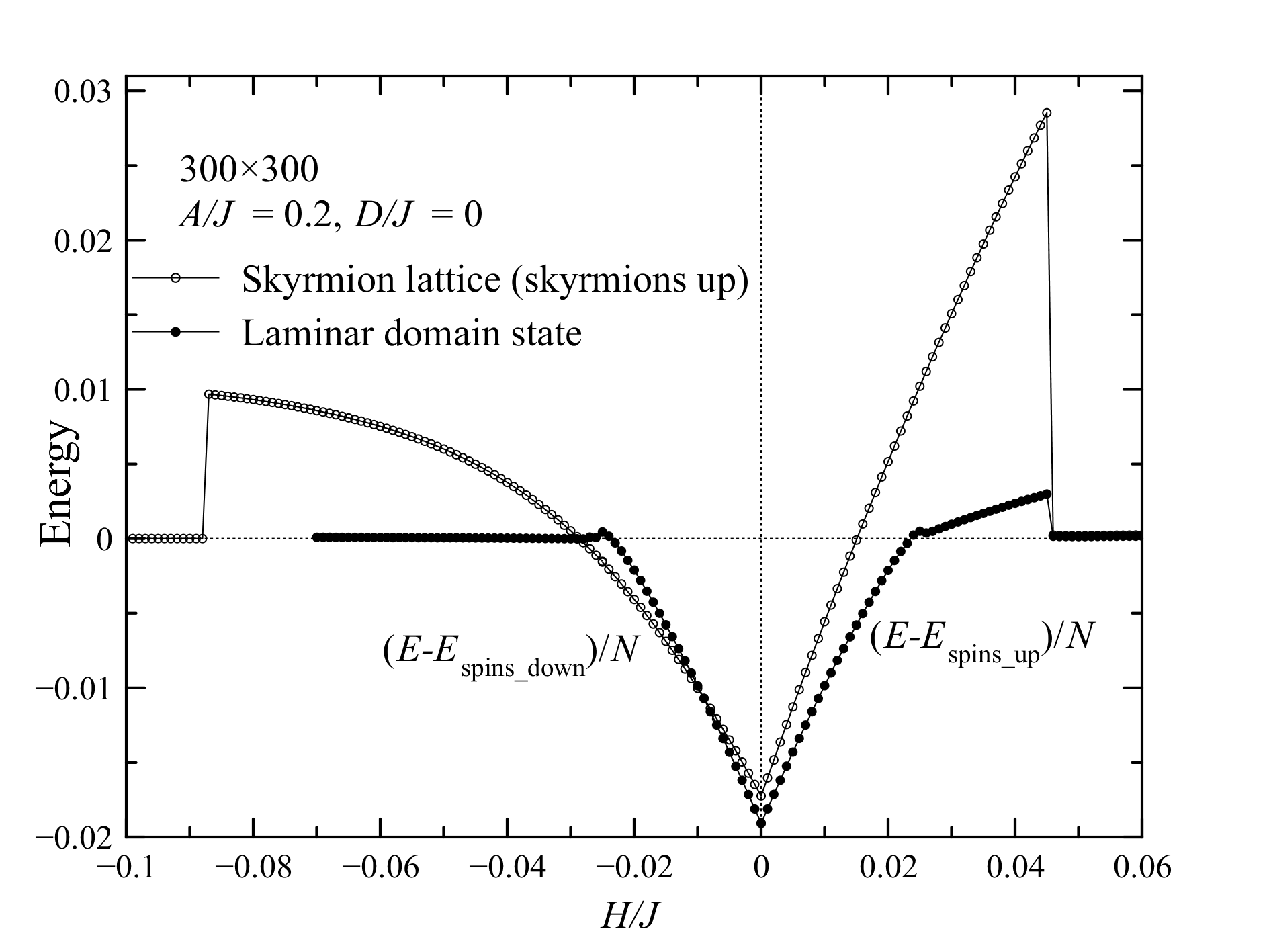} \includegraphics[width=8cm]{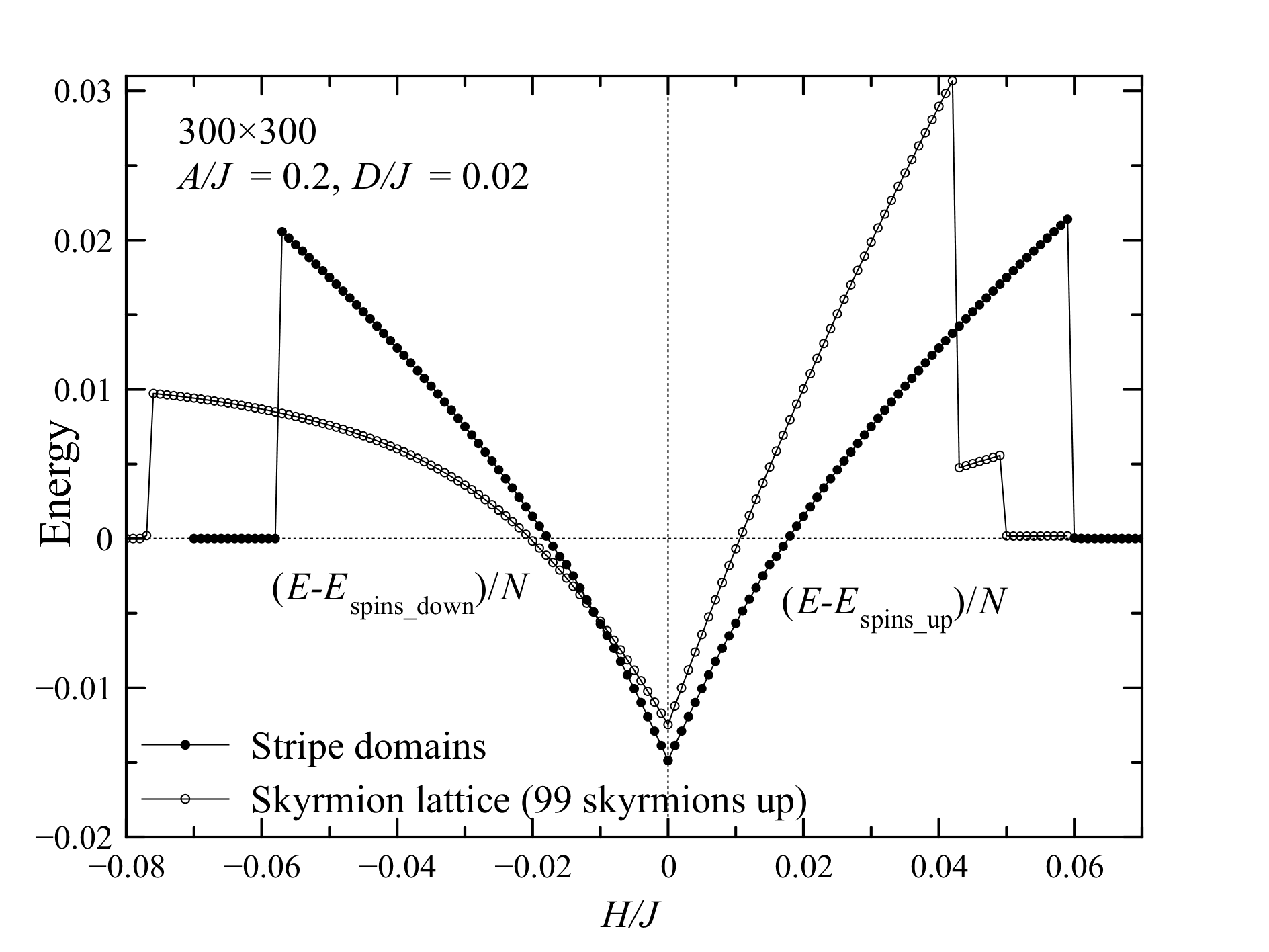}
\par\end{centering}
\caption{Energy of the skyrmion lattice (skyrmions pointing up against the
spin-down background) and the domain state with respect to the energy
of the uniformly magnetized state with the spins up or down. The skyrmion
lattice is the ground state in a narrow region of negative fields.
Upper panel: $D=0$. Lower panel: $D/J=0.02$. }
\label{energy}
\end{figure}
Next we compare energies of the skyrmion and domain states. In Fig.\ \ref{energy}
they are plotted as function of the field with respect to the energy of the uniformly magnetized
state with spins directed up or down. The
skyrmion lattice has the energy lower than the domain structure and
lower than the uniformly magnetized state in a narrow range of negative
fields that becomes more narrow with increasing PMA.

\section{Discussion}

\label{Sec_Discussion} We have studied theoretically, via Monte Carlo
simulations on large spin lattices, the skyrmion states that emerge along the hysteresis curve in a ferromagnetic
film with the DMI, PMA, and RMA  at zero and
finite temperatures. Our results agree with recent experimental findings
\cite{Zhang2018} on thermally assisted creation of skyrmions, thus
providing theoretical background for these findings.

In accordance with experiments, we find two mechanisms of the creation
of skyrmions: 1) via thermally assisted nucleation from the uniformly
magnetized state and 2) via destruction of labyrinth magnetic domains
into skyrmions. These skyrmion states appear in positive and negative
fields respectively along the magnetization curve. In Ref.\ \onlinecite{Zhang2018},
the temperature dependence of the PMA has been independently measured
and the correlation of the amplitude of the PMA with the field range
where skyrmions exist has been found. 

In our studies we choose the
zero-temperature value of the PMA as a microscopic parameter and 
obtain the decrease of the effective PMA with temperature self-consistently, owed to thermal
fluctuations of the spins. In agreement with experiment,
we find that in a system without quenched randomness the skyrmion
states exist in a narrow field range suppressed by the PMA.

At $T=0$ conservation of the topological charge suppresses creation
of skyrmions unless static randomness is present in the system. We
model such randomness by the RMA. In contrast with the PMA, the RMA,
similarly to the temperature but not as effective, creates a non-zero topological charge in a broad field
range even at $T=0$. While our results agree qualitatively with a large body of previous work on slyrmion states, see, e.g., Ref.\ \onlinecite{Leonov-NJP2016}, the account of the temperature dependence of the magnetic anisotropy and of the effect of random magnetic anisotropy provide important new nuances to this problem.  

We have observed in numerical experiments how skyrmions merge into
labyrinth domains and how the domains decay back into skyrmions on
changing the field. Properties of skyrmion states depend strongly
on the manner in which they are obtained. We have studied skyrmion
lattices generated by changing the field along the hysteresis curve, formation of skyrmion structures
assisted by static randomness, by temperature, by freezing the system
from elevated temperature to $T=0$, as well as skyrmion lattices
emerging from random initial conditions for the spins.

The work on skyrmionics, that began at low temperatures, has now moved to room temperatures and above. This has made possible the experiments
on thermally assisted creation of skyrmions, like the ones described
in Ref.\ \onlinecite{Zhang2018}. We hope that the findings reported
in this paper will further assist planning and understanding of such experiments.

\section{Acknowledgements}

This work has been supported by the grant No. OSR-2016-CRG5-2977 from
King Abdullah University of Science and Technology.

\end{document}